\documentclass[dvips]{article}
\usepackage{supertabular,lscape,epsfig}
\usepackage{amssymb}
\usepackage{amsmath}

\usepackage[polish]{babel}
\usepackage[T1]{fontenc}
\usepackage[latin2]{inputenc}
\usepackage{pslatex}

\DeclareSymbolFont{ppa}{OT1}{ppl}{m}{it}
\DeclareMathSymbol{\vv}{\mathalpha}{ppa}{'166}

\begin{document}

\newcommand{\dd}{\,{\rm d}}
\newcommand{\ie}{{\it i.e.},\,}
\newcommand{\etal}{{\it et al.\ }}
\newcommand{\eg}{{\it e.g.},\,}
\newcommand{\cf}{{\it cf.\ }}
\newcommand{\vs}{{\it vs.\ }}
\newcommand{\zdot}{\makebox[0pt][l]{.}}
\newcommand{\up}[1]{\ifmmode^{\rm #1}\else$^{\rm #1}$\fi}
\newcommand{\dn}[1]{\ifmmode_{\rm #1}\else$_{\rm #1}$\fi}
\newcommand{\upd}{\up{d}}
\newcommand{\uph}{\up{h}}
\newcommand{\upm}{\up{m}}  
\newcommand{\ups}{\up{s}}
\newcommand{\arcd}{\ifmmode^{\circ}\else$^{\circ}$\fi}
\newcommand{\arcm}{\ifmmode{'}\else$'$\fi}
\newcommand{\arcs}{\ifmmode{''}\else$''$\fi}
\newcommand{\MS}{{\rm M}\ifmmode_{\odot}\else$_{\odot}$\fi}
\newcommand{\RS}{{\rm R}\ifmmode_{\odot}\else$_{\odot}$\fi}
\newcommand{\LS}{{\rm L}\ifmmode_{\odot}\else$_{\odot}$\fi}

\newcommand{\Abstract}[2]{{\footnotesize\begin{center}ABSTRACT\end{center}
\vspace{1mm}\par#1\par   
\noindent
{~}{\it #2}}}

\newcommand{\TabCap}[2]{\begin{center}\parbox[t]{#1}{\begin{center}
  \small {\spaceskip 2pt plus 1pt minus 1pt T a b l e}
  \refstepcounter{table}\thetable \\[2mm]
  \footnotesize #2 \end{center}}\end{center}}

\newcommand{\TableSep}[2]{\begin{table}[p]\vspace{#1}
\TabCap{#2}\end{table}}

\newcommand{\FigCap}[1]{\footnotesize\par\noindent Fig.\  %
  \refstepcounter{figure}\thefigure. #1\par}

\newcommand{\TableFont}{\footnotesize}
\newcommand{\TableFontIt}{\ttit}
\newcommand{\SetTableFont}[1]{\renewcommand{\TableFont}{#1}}

\newcommand{\MakeTable}[4]{\begin{table}[htb]\TabCap{#2}{#3}
  \begin{center} \TableFont \begin{tabular}{#1} #4
  \end{tabular}\end{center}\end{table}}

\newcommand{\MakeTableSep}[4]{\begin{table}[p]\TabCap{#2}{#3}
  \begin{center} \TableFont \begin{tabular}{#1} #4
  \end{tabular}\end{center}\end{table}}

\newenvironment{references}%
{
\footnotesize \frenchspacing
\renewcommand{\thesection}{}
\renewcommand{\in}{{\rm in }}
\renewcommand{\AA}{Astron.\ Astrophys.}
\newcommand{\AAS}{Astron.~Astrophys.~Suppl.~Ser.}
\newcommand{\ApJ}{Astrophys.\ J.}
\newcommand{\ApJS}{Astrophys.\ J.~Suppl.~Ser.}
\newcommand{\ApJL}{Astrophys.\ J.~Letters}
\newcommand{\AJ}{Astron.\ J.}
\newcommand{\IBVS}{IBVS}
\newcommand{\PASP}{P.A.S.P.}
\newcommand{\Acta}{Acta Astron.}
\newcommand{\MNRAS}{MNRAS}
\renewcommand{\and}{{\rm and }}
\section{{\rm REFERENCES}}
\sloppy \hyphenpenalty10000
\begin{list}{}{\leftmargin1cm\listparindent-1cm
\itemindent\listparindent\parsep0pt\itemsep0pt}}%
{\end{list}\vspace{2mm}}
 
\def\TYLDA{~}
\newlength{\DW}
\settowidth{\DW}{0}
\newcommand{\dw}{\hspace{\DW}}

\newcommand{\refitem}[5]{\item[]{#1} #2%
\def\REFARG{#3}\ifx\REFARG\TYLDA\else, {\it#3}\fi
\def\REFARG{#4}\ifx\REFARG\TYLDA\else, {\bf#4}\fi
\def\REFARG{#5}\ifx\REFARG\TYLDA\else, {#5}\fi.}

\newcommand{\Section}[1]{\section{#1}}
\newcommand{\Subsection}[1]{\subsection{#1}}
\newcommand{\Acknow}[1]{\par\vspace{5mm}{\bf Acknowledgements.} #1}
\pagestyle{myheadings}

\newfont{\bb}{ptmbi8t at 12pt}
\newcommand{\xrule}{\rule{0pt}{2.5ex}}  
\newcommand{\xxrule}{\rule[-1.8ex]{0pt}{4.5ex}}  
\def\thefootnote{\fnsymbol{footnote}}
\begin{center}

{\Large\bf
The Optical Gravitational Lensing Experiment.\\
\vskip2pt
OGLE-III Photometric Maps of the Large Magellanic Cloud\footnote{Based on
observations obtained with the 1.3 m Warsaw telescope at the Las Campanas
Observatory of the Carnegie Institution of Washington.}}
\vskip1.5cm
{\bf A.~~U~d~a~l~s~k~i$^1$,~~ I.~~S~o~s~z~y~ń~s~k~i$^1$,~~ 
M.\,K.~~ S~z~y~m~a~ń~s~k~i$^1$,~~ M.~~ K~u~b~i~a~k$^1$,\\
G.~~ P~i~e~t~r~z~y~ń~s~k~i$^{1,2}$,~~
Ł.~~ W~y~r~z~y~k~o~w~s~k~i$^{3,1}$,~~
O.~~ S~z~e~w~c~z~y~k$^{2,1}$,\\
K.~~ U~l~a~c~z~y~k$^1$~~
and~~ R.~~P~o~l~e~s~k~i$^1$}
\vskip5mm
$^1$Warsaw University Observatory, Al. Ujazdowskie~4, 
00-478~Warszawa, Poland\\
e-mail:
(udalski,soszynsk,msz,mk,pietrzyn,wyrzykow,szewczyk,kulaczyk,rpoleski)
@astrouw.edu.pl\\
$^2$ Universidad de Concepci{\'o}n, Departamento de Fisica, Casilla 160--C,
Concepci{\'o}n, Chile\\ 
$^3$ Institute of Astronomy, University of Cambridge, Madingley Road,
Cambridge CB3~0HA,~UK
\end{center}

\Abstract{
We present the OGLE-III Photometric Maps of the Large Magellanic Cloud.
They cover about 40 square degrees of the LMC and contain mean, calibrated
{\it VI} photometry and astrometry of about 35 million stars observed
during seven observing seasons of the third phase of the Optical
Gravitational Lensing Experiment -- OGLE-III.

We discuss the quality of data and present color--magnitude diagrams of
selected fields. The OGLE-III Photometric Maps of the LMC are available to
the astronomical community from the OGLE Internet archive.}{Magellanic
Clouds -- Surveys -- Catalogs -- Techniques: photometric}

\Section{Introduction}
One of the important results of the second phase of the Optical
Gravitational Lensing Experiment was publication of the OGLE Photometric
Maps of Dense Stellar Regions (Udalski \etal 1998, 2000, 2002) containing
precisely calibrated {\it BVI} photometry and astrometry of the SMC, LMC and
Galactic bulge fields observed during OGLE-II. Because these regions of the
sky are extremely interesting from the astrophysical point of view, OGLE
maps have been widely used by astronomers worldwide for many astrophysical
applications.
   
OGLE-III phase of the OGLE project, started on June 12, 2001 and still
in operation, was a significant extension of the OGLE survey. Much larger
observing capabilities made it possible to cover practically entire area
of the LMC and SMC and large fraction of the Galactic bulge. After seven
years of continuous observations the huge collection of OGLE images was
re-reduced to obtain the final precise photometry, calibrated to the
standard system (Udalski \etal 2008). Thus it became possible to extend
the OGLE-II maps to new regions, not observed by OGLE before.

In this paper we present the OGLE-III photometric maps of the Large
Magellanic Cloud. They are available to the astronomical community from
the OGLE Internet archive.

\Section{Observations}
OGLE-III images of the LMC used for construction of the OGLE-III
Photometric Maps of this galaxy were collected between July 2001 and
March 2008 and cover seven observing seasons of the LMC. Observations
were carried out at Las Campanas Observatory, operated by the Carnegie
Institution of Washington, with 1.3~m Warsaw telescope equipped with the
eight chip mosaic camera (Udalski 2003). One full mosaic image covers
approximately $35\arcm\times35\arcm$ on the sky with the scale of
0\zdot\arcs26/pixel.

As the main goal of the OGLE survey is the search for variability of
observed objects, the majority of observations were obtained through the
single, namely {\it I}-band filter. Although this filter well approximates
the standard one for $V-I<3$~mag colors, one has to be aware that for very
red objects some deviation may be present. Several hundred {\it I}-band
images were collected for each of the observed fields. From time to time
the LMC fields were also observed in the {\it V}-band -- typically about
40--50 times in the considered period. The exposure time was set to 180 and
240 seconds in the {\it I} and {\it V}-band, respectively.

Because the majority of observed fields have high stellar density,
observations were conducted only in good seeing and transparency
conditions. When the seeing exceeded 1\zdot\arcs8 observations were
stopped. The median seeing of the {\it V} and {\it I}-band OGLE-III LMC
datasets is equal to 1\zdot\arcs2.

Table~1 lists all LMC fields observed during OGLE-III phase. It also
provides equatorial coordinates of their centers and number of stars
detected in the {\it I}-band. The total observed area reaches 40 square
degrees. Fig.~1 presents a combined image of the LMC taken by the ASAS
survey program (Pojmański, 1997) with contours of the OGLE-II and
OGLE-III  fields.

\Section{Data Reductions} 
Initial pre-reductions of the collected images -- de-biasing, flatfielding
-- are done at the telescope, immediately after the last pixel of the image
is written to the main data acquisition computer (Udalski 2003). Although
the first, provisional photometry is done also in almost real time at the
telescope, photometry used in this paper comes from the final off-line
re-reductions of the entire LMC dataset collected so far. Photometry is
based on the Difference Image Analysis method (DIA -- Alard and Lupton
1998, Woźniak 2000) and all details on the OGLE-III implementation and
calibration to the standard system can be found in Udalski \etal
(2008). Comparison with the OGLE-II photometric maps indicates that the
mean difference of the magnitudes between the calibrated OGLE-III and
OGLE-II photometry for about 800000 stars brighter than $I<18$~mag and
$V<19$~mag in overlaping fields is negligible ($-0.004\pm0.013,
-0.003\pm0.013$ in the {\it I} and {\it V}, respectively when calibrating
directly with standard stars from Landolt's (1992) fields or
$0.001\pm0.010, 0.000\pm0.007$ for final calibrations of OGLE-III
photometry with OGLE-II maps) implying that OGLE-II and OGLE-III maps are
photometrically fully consistent.

Astrometric transformation of the pixel grid to equatorial system was done
in similar way as in the OGLE-II maps. Details can also be found in Udalski
\etal (2008).

\renewcommand{\TableFont}{\scriptsize}
\renewcommand{\arraystretch}{0.7}
\MakeTableSep{cccrccccr}{12.5cm}{OGLE-III Fields in the LMC}
{\cline{1-4}\cline{6-9}
\noalign{\vskip3pt}
Field & RA       &   DEC   & $N_{\rm Stars}$ &$\phantom{xxxxx}$& Field & RA       &   DEC   & $N_{\rm Stars}$ \\
    & (2000)   &  (2000) &                 &&     & (2000)   &  (2000) &                \\
\noalign{\vskip3pt}
\cline{1-4}\cline{6-9}
\noalign{\vskip3pt}
LMC100 & 5\uph19\upm02\zdot\ups2 & $-69\arcd15\arcm07\arcs$ &  1075711 && LMC158 & 4\uph30\upm59\zdot\ups9 & $-70\arcd26\arcm01\arcs$ &    34264\\
LMC101 & 5\uph19\upm03\zdot\ups1 & $-68\arcd39\arcm19\arcs$ &   548707 && LMC159 & 5\uph25\upm11\zdot\ups4 & $-68\arcd03\arcm58\arcs$ &   234254\\
LMC102 & 5\uph19\upm03\zdot\ups4 & $-68\arcd03\arcm48\arcs$ &   266963 && LMC160 & 5\uph25\upm20\zdot\ups9 & $-68\arcd39\arcm24\arcs$ &   437897\\
LMC103 & 5\uph19\upm02\zdot\ups9 & $-69\arcd50\arcm26\arcs$ &   833034 && LMC161 & 5\uph25\upm32\zdot\ups5 & $-69\arcd14\arcm59\arcs$ &   787467\\
LMC104 & 5\uph19\upm02\zdot\ups4 & $-70\arcd26\arcm03\arcs$ &   472959 && LMC162 & 5\uph25\upm43\zdot\ups3 & $-69\arcd50\arcm24\arcs$ &  1171172\\
LMC105 & 5\uph19\upm01\zdot\ups6 & $-71\arcd01\arcm31\arcs$ &   353469 && LMC163 & 5\uph25\upm52\zdot\ups2 & $-70\arcd25\arcm55\arcs$ &   806912\\
LMC106 & 5\uph19\upm01\zdot\ups0 & $-71\arcd36\arcm57\arcs$ &   236384 && LMC164 & 5\uph26\upm08\zdot\ups4 & $-71\arcd01\arcm23\arcs$ &   381352\\
LMC107 & 5\uph13\upm01\zdot\ups5 & $-66\arcd52\arcm57\arcs$ &   226287 && LMC165 & 5\uph26\upm20\zdot\ups9 & $-71\arcd37\arcm01\arcs$ &   329028\\
LMC108 & 5\uph13\upm01\zdot\ups9 & $-67\arcd28\arcm40\arcs$ &   257801 && LMC166 & 5\uph31\upm20\zdot\ups1 & $-68\arcd03\arcm51\arcs$ &   314715\\
LMC109 & 5\uph12\upm53\zdot\ups3 & $-68\arcd04\arcm06\arcs$ &   294145 && LMC167 & 5\uph31\upm39\zdot\ups6 & $-68\arcd39\arcm32\arcs$ &   394585\\
LMC110 & 5\uph12\upm43\zdot\ups6 & $-68\arcd39\arcm42\arcs$ &   529605 && LMC168 & 5\uph32\upm01\zdot\ups4 & $-69\arcd15\arcm00\arcs$ &   636559\\
LMC111 & 5\uph12\upm32\zdot\ups7 & $-69\arcd15\arcm02\arcs$ &   696121 && LMC169 & 5\uph32\upm22\zdot\ups8 & $-69\arcd50\arcm26\arcs$ &  1092986\\
LMC112 & 5\uph12\upm21\zdot\ups5 & $-69\arcd50\arcm21\arcs$ &   752907 && LMC170 & 5\uph32\upm48\zdot\ups1 & $-70\arcd25\arcm53\arcs$ &   878425\\
LMC113 & 5\uph12\upm10\zdot\ups9 & $-70\arcd25\arcm48\arcs$ &   561679 && LMC171 & 5\uph33\upm10\zdot\ups6 & $-71\arcd01\arcm30\arcs$ &   513519\\
LMC114 & 5\uph11\upm58\zdot\ups9 & $-71\arcd01\arcm22\arcs$ &   221912 && LMC172 & 5\uph33\upm34\zdot\ups4 & $-71\arcd36\arcm54\arcs$ &   423099\\
LMC115 & 5\uph07\upm09\zdot\ups7 & $-66\arcd52\arcm59\arcs$ &   234177 && LMC173 & 5\uph37\upm29\zdot\ups3 & $-68\arcd03\arcm50\arcs$ &   221971\\
LMC116 & 5\uph07\upm00\zdot\ups9 & $-67\arcd28\arcm29\arcs$ &   205536 && LMC174 & 5\uph37\upm59\zdot\ups8 & $-68\arcd39\arcm26\arcs$ &   340208\\
LMC117 & 5\uph06\upm55\zdot\ups3 & $-68\arcd03\arcm58\arcs$ &   527346 && LMC175 & 5\uph38\upm32\zdot\ups3 & $-69\arcd15\arcm01\arcs$ &   497977\\
LMC118 & 5\uph06\upm25\zdot\ups4 & $-68\arcd39\arcm25\arcs$ &   694697 && LMC176 & 5\uph39\upm01\zdot\ups6 & $-69\arcd50\arcm30\arcs$ &   576984\\
LMC119 & 5\uph06\upm02\zdot\ups5 & $-69\arcd15\arcm02\arcs$ &   817851 && LMC177 & 5\uph39\upm38\zdot\ups0 & $-70\arcd25\arcm49\arcs$ &   817966\\
LMC120 & 5\uph05\upm39\zdot\ups8 & $-69\arcd50\arcm28\arcs$ &   617701 && LMC178 & 5\uph40\upm14\zdot\ups1 & $-71\arcd01\arcm27\arcs$ &   471477\\
LMC121 & 5\uph05\upm14\zdot\ups4 & $-70\arcd25\arcm59\arcs$ &   442352 && LMC179 & 5\uph40\upm52\zdot\ups3 & $-71\arcd36\arcm58\arcs$ &   296759\\
LMC122 & 5\uph04\upm52\zdot\ups9 & $-71\arcd01\arcm25\arcs$ &   288282 && LMC180 & 5\uph40\upm51\zdot\ups5 & $-72\arcd12\arcm28\arcs$ &   258973\\
LMC123 & 5\uph01\upm18\zdot\ups0 & $-66\arcd53\arcm00\arcs$ &   242827 && LMC181 & 5\uph43\upm35\zdot\ups7 & $-68\arcd03\arcm58\arcs$ &   201579\\
LMC124 & 5\uph01\upm00\zdot\ups3 & $-67\arcd28\arcm27\arcs$ &   290411 && LMC182 & 5\uph44\upm16\zdot\ups0 & $-68\arcd39\arcm32\arcs$ &   311362\\
LMC125 & 5\uph00\upm36\zdot\ups1 & $-68\arcd03\arcm54\arcs$ &   357288 && LMC183 & 5\uph45\upm02\zdot\ups8 & $-69\arcd14\arcm59\arcs$ &   361838\\
LMC126 & 5\uph00\upm02\zdot\ups4 & $-68\arcd39\arcm31\arcs$ &   530735 && LMC184 & 5\uph45\upm43\zdot\ups2 & $-69\arcd50\arcm33\arcs$ &   486666\\
LMC127 & 4\uph59\upm33\zdot\ups6 & $-69\arcd14\arcm54\arcs$ &   547901 && LMC185 & 5\uph46\upm30\zdot\ups8 & $-70\arcd25\arcm51\arcs$ &   630366\\
LMC128 & 4\uph59\upm03\zdot\ups6 & $-69\arcd50\arcm24\arcs$ &   406243 && LMC186 & 5\uph47\upm21\zdot\ups2 & $-71\arcd01\arcm24\arcs$ &   376966\\
LMC129 & 4\uph58\upm24\zdot\ups6 & $-70\arcd26\arcm07\arcs$ &   304616 && LMC187 & 5\uph48\upm12\zdot\ups6 & $-71\arcd36\arcm52\arcs$ &   298070\\
LMC130 & 4\uph57\upm50\zdot\ups8 & $-71\arcd01\arcm20\arcs$ &   231039 && LMC188 & 5\uph48\upm26\zdot\ups6 & $-72\arcd12\arcm27\arcs$ &   152482\\
LMC131 & 4\uph55\upm28\zdot\ups6 & $-66\arcd52\arcm46\arcs$ &   248421 && LMC189 & 5\uph50\upm37\zdot\ups9 & $-68\arcd39\arcm26\arcs$ &   153656\\
LMC132 & 4\uph55\upm00\zdot\ups6 & $-67\arcd28\arcm36\arcs$ &   217827 && LMC190 & 5\uph51\upm33\zdot\ups2 & $-69\arcd14\arcm55\arcs$ &   200564\\
LMC133 & 4\uph54\upm29\zdot\ups2 & $-68\arcd03\arcm47\arcs$ &   346558 && LMC191 & 5\uph52\upm20\zdot\ups1 & $-69\arcd50\arcm24\arcs$ &   243558\\
LMC134 & 4\uph53\upm49\zdot\ups2 & $-68\arcd39\arcm18\arcs$ &   304271 && LMC192 & 5\uph53\upm24\zdot\ups1 & $-70\arcd25\arcm51\arcs$ &   228516\\
LMC135 & 4\uph53\upm05\zdot\ups2 & $-69\arcd14\arcm51\arcs$ &   284846 && LMC193 & 5\uph54\upm21\zdot\ups7 & $-71\arcd01\arcm34\arcs$ &   138223\\
LMC136 & 4\uph52\upm23\zdot\ups7 & $-69\arcd50\arcm25\arcs$ &   244490 && LMC194 & 5\uph55\upm29\zdot\ups7 & $-71\arcd36\arcm59\arcs$ &    77816\\
LMC137 & 4\uph51\upm30\zdot\ups2 & $-70\arcd26\arcm01\arcs$ &   200058 && LMC195 & 5\uph56\upm00\zdot\ups0 & $-72\arcd12\arcm25\arcs$ &    43002\\
LMC138 & 4\uph49\upm34\zdot\ups7 & $-66\arcd53\arcm07\arcs$ &   117664 && LMC196 & 5\uph56\upm54\zdot\ups7 & $-68\arcd39\arcm29\arcs$ &   116258\\
LMC139 & 4\uph49\upm05\zdot\ups2 & $-67\arcd28\arcm30\arcs$ &   128858 && LMC197 & 5\uph58\upm02\zdot\ups7 & $-69\arcd15\arcm06\arcs$ &    92295\\
LMC140 & 4\uph48\upm18\zdot\ups2 & $-68\arcd04\arcm05\arcs$ &   188164 && LMC198 & 5\uph59\upm02\zdot\ups5 & $-69\arcd50\arcm35\arcs$ &    71992\\
LMC141 & 4\uph47\upm26\zdot\ups7 & $-68\arcd39\arcm36\arcs$ &   191197 && LMC199 & 6\uph00\upm14\zdot\ups7 & $-70\arcd26\arcm00\arcs$ &    63841\\
LMC142 & 4\uph46\upm31\zdot\ups9 & $-69\arcd15\arcm08\arcs$ &   225807 && LMC200 & 6\uph01\upm27\zdot\ups5 & $-71\arcd01\arcm36\arcs$ &    58765\\
LMC143 & 4\uph45\upm43\zdot\ups1 & $-69\arcd50\arcm19\arcs$ &   155672 && LMC201 & 6\uph02\upm45\zdot\ups9 & $-71\arcd37\arcm04\arcs$ &    91051\\
LMC144 & 4\uph44\upm40\zdot\ups2 & $-70\arcd26\arcm01\arcs$ &   116363 && LMC202 & 6\uph03\upm28\zdot\ups3 & $-72\arcd12\arcm34\arcs$ &    79469\\
LMC145 & 4\uph43\upm47\zdot\ups5 & $-66\arcd52\arcm43\arcs$ &    64628 && LMC203 & 6\uph03\upm29\zdot\ups9 & $-72\arcd48\arcm04\arcs$ &    71396\\
LMC146 & 4\uph43\upm03\zdot\ups0 & $-67\arcd28\arcm17\arcs$ &    86656 && LMC204 & 6\uph03\upm14\zdot\ups6 & $-68\arcd39\arcm25\arcs$ &   108173\\
LMC147 & 4\uph42\upm07\zdot\ups8 & $-68\arcd03\arcm55\arcs$ &   103604 && LMC205 & 6\uph04\upm32\zdot\ups9 & $-69\arcd15\arcm04\arcs$ &    82072\\
LMC148 & 4\uph41\upm06\zdot\ups8 & $-68\arcd39\arcm27\arcs$ &   110885 && LMC206 & 6\uph05\upm40\zdot\ups3 & $-69\arcd50\arcm27\arcs$ &    78179\\
LMC149 & 4\uph40\upm05\zdot\ups1 & $-69\arcd14\arcm57\arcs$ &   115117 && LMC207 & 6\uph07\upm04\zdot\ups2 & $-70\arcd25\arcm55\arcs$ &    70889\\
LMC150 & 4\uph39\upm05\zdot\ups3 & $-69\arcd50\arcm16\arcs$ &    96039 && LMC208 & 6\uph08\upm30\zdot\ups4 & $-71\arcd01\arcm27\arcs$ &    87619\\
LMC151 & 4\uph37\upm51\zdot\ups6 & $-70\arcd25\arcm45\arcs$ &    89935 && LMC209 & 6\uph10\upm07\zdot\ups0 & $-71\arcd37\arcm00\arcs$ &    64885\\
LMC152 & 4\uph37\upm54\zdot\ups1 & $-66\arcd52\arcm52\arcs$ &    46107 && LMC210 & 6\uph10\upm55\zdot\ups7 & $-72\arcd12\arcm37\arcs$ &    70282\\
LMC153 & 4\uph37\upm01\zdot\ups7 & $-67\arcd28\arcm30\arcs$ &    56016 && LMC211 & 6\uph11\upm22\zdot\ups0 & $-72\arcd48\arcm04\arcs$ &    61205\\
LMC154 & 4\uph35\upm59\zdot\ups1 & $-68\arcd04\arcm02\arcs$ &    65095 && LMC212 & 6\uph11\upm04\zdot\ups0 & $-69\arcd14\arcm50\arcs$ &    81878\\
LMC155 & 4\uph34\upm49\zdot\ups4 & $-68\arcd39\arcm32\arcs$ &    71173 && LMC213 & 6\uph12\upm17\zdot\ups9 & $-69\arcd50\arcm37\arcs$ &    52207\\
LMC156 & 4\uph33\upm32\zdot\ups7 & $-69\arcd15\arcm00\arcs$ &    72805 && LMC214 & 6\uph13\upm58\zdot\ups2 & $-70\arcd26\arcm08\arcs$ &    60410\\
LMC157 & 4\uph32\upm23\zdot\ups8 & $-69\arcd50\arcm26\arcs$ &    44165 && LMC215 & 6\uph15\upm36\zdot\ups4 & $-71\arcd01\arcm28\arcs$ &    60801\\
\noalign{\vskip3pt}
\cline{1-4}\cline{6-9}}

\Section{Construction of Photometric Maps}
Because the OGLE-III photometric databases are constructed separately
for {\it I} and {\it V}-band data, the first step of map preparation was
the cross-identification of {\it V}-band counterparts to each {\it
I}-band object. Due to some shifts between respective {\it I} and {\it
V}-band reference images of a given subfield and larger than single
detector dimensions size of the reference images to mask the gaps
between chips of the OGLE-III mosaic camera, the {\it V}-band
counterparts of objects located close to the border in the {\it I}-band
reference image may be present in more than one field. Therefore in the
first step third order transformation between pixel grids of a given
{\it I}-band subfield reference image and all {\it V}-band reference
images of subfields that even partially overlap were derived. Then for
all {\it I}-band database objects of a given subfield their
corresponding {\it V}-band counterparts in all fields were found.

The mean photometry was derived for all objects with minimum of 4 and 6
observations in the {\it V} and {\it I}-band, respectively, by averaging
all observations after removing $5\sigma$ deviating points. In the case
of multiple {\it V}-band cross-identifications all {\it V}-band
observations formed a single dataset and then were averaged because they
are independent measurements. Finally, the color term correction was
applied for each object to its database average magnitude according to
the transformation equations and color term coefficients presented in
Udalski \etal (2008). After the transformation of the $V-I$ color to the
standard system, {\it I} and {\it V} magnitudes were appropriately
adjusted. For objects that do not have color information the average
color of the LMC population $V-I=0.7$~mag was used for color correction
of respective {\it I} or {\it V} magnitude.

Table~2 presents the first 25 entries from the map of the LMC100.1
subfield. The columns contain: (1) ID number; (2,3) equatorial coordinates
J2000.0; (4,5) $X,Y$ pixel coordinates in the {\it I}-band reference image;
(6,7,8) photometry: {\it V}, $V-I$, {\it I}; (9,10,11) number of points for
average magnitude in {\it V}, number of $5\sigma$ removed points in {\it
V}, $\sigma$ of magnitude for {\it V}-band; (12,13,14) same as (9,10,11)
for the {\it I}-band. 9.999 or 99.999 markers mean ``no data''. $-1$ in
column (10) indicates multiple {\it V}-band cross-identification (the
average magnitude and standard deviation are calculated for merged
photometry).

The full set of the OGLE-III Photometric Maps is available from the OGLE
Internet archive (see below).

\Section{Discussion}
OGLE-III Photometric Maps form a significant extension of the OGLE-II maps
that covered only central regions of the LMC (\cf Fig.~1). The new maps
include practically entire area of the LMC and can be used for many
projects studying the global structure of this galaxy. Precise, well
calibrated {\it VI} photometry and astrometry makes this dataset a
unique tool for many astrophysical applications.

To show the accuracy of the OGLE-III Photometric Maps Figs.~2 and 3 present
standard deviation of magnitudes as a function of magnitude in the {\it V}
and {\it I}-band for two LMC fields: LMC100.1 -- very dense field located
in the central bar and LMC209.1 -- sparsely populated field form the outer
parts of this galaxy. As one can expect the accuracy of photometry depends
on the stellar density and, for example, $\sigma=0.1$~mag photometry
scatter is reached for stars by about 0.3 magnitude brighter in the densest
bar fields than in the uncrowded fields.

\setcounter{figure}{3}
\begin{figure}[p]
\centerline{\includegraphics[width=9cm, bb=30 50 510 550]{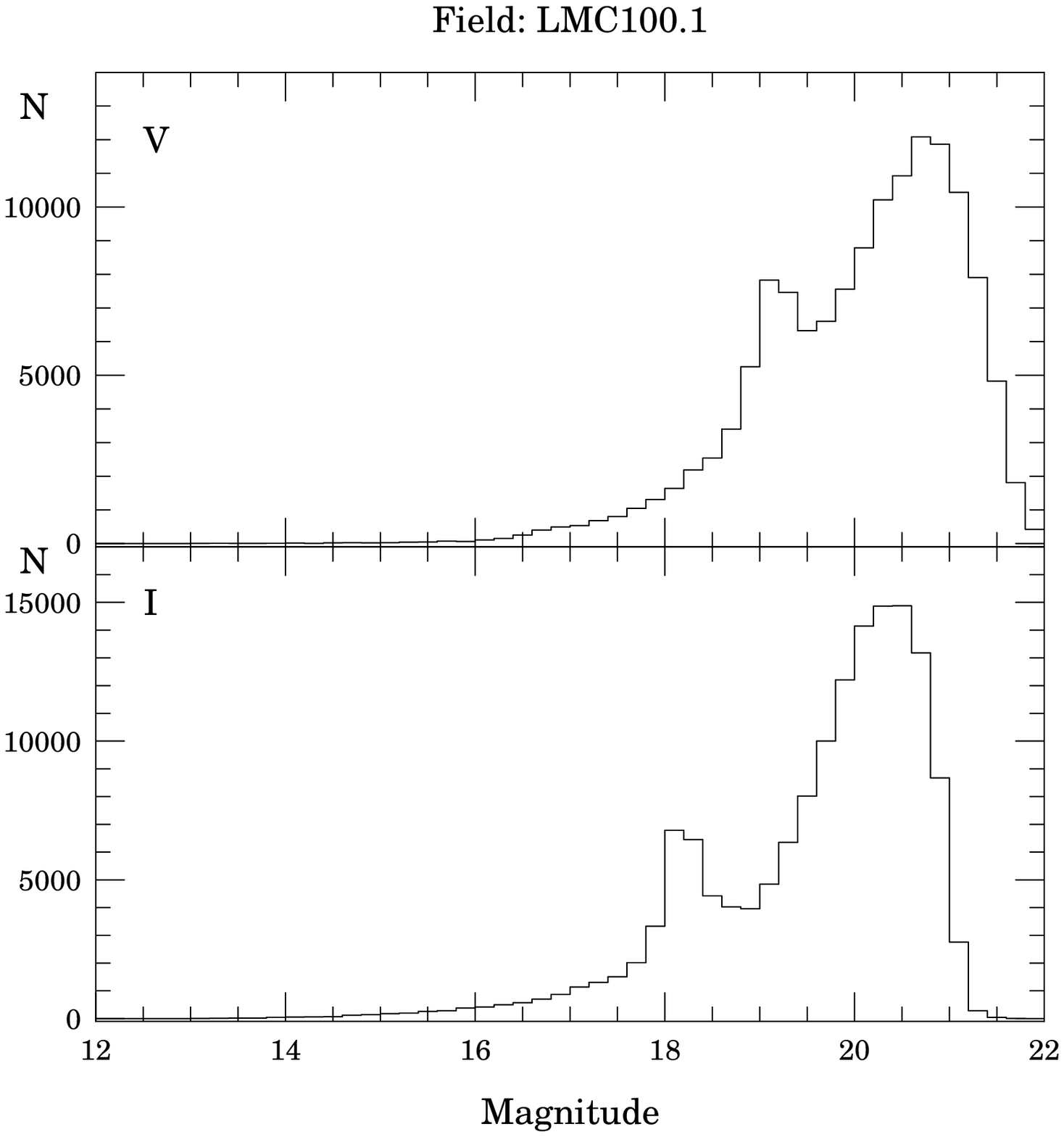}}
\FigCap{Histogram of magnitudes for the central bar subfield LMC100.1.}
\centerline{\includegraphics[width=9cm, bb=30 50 510 550]{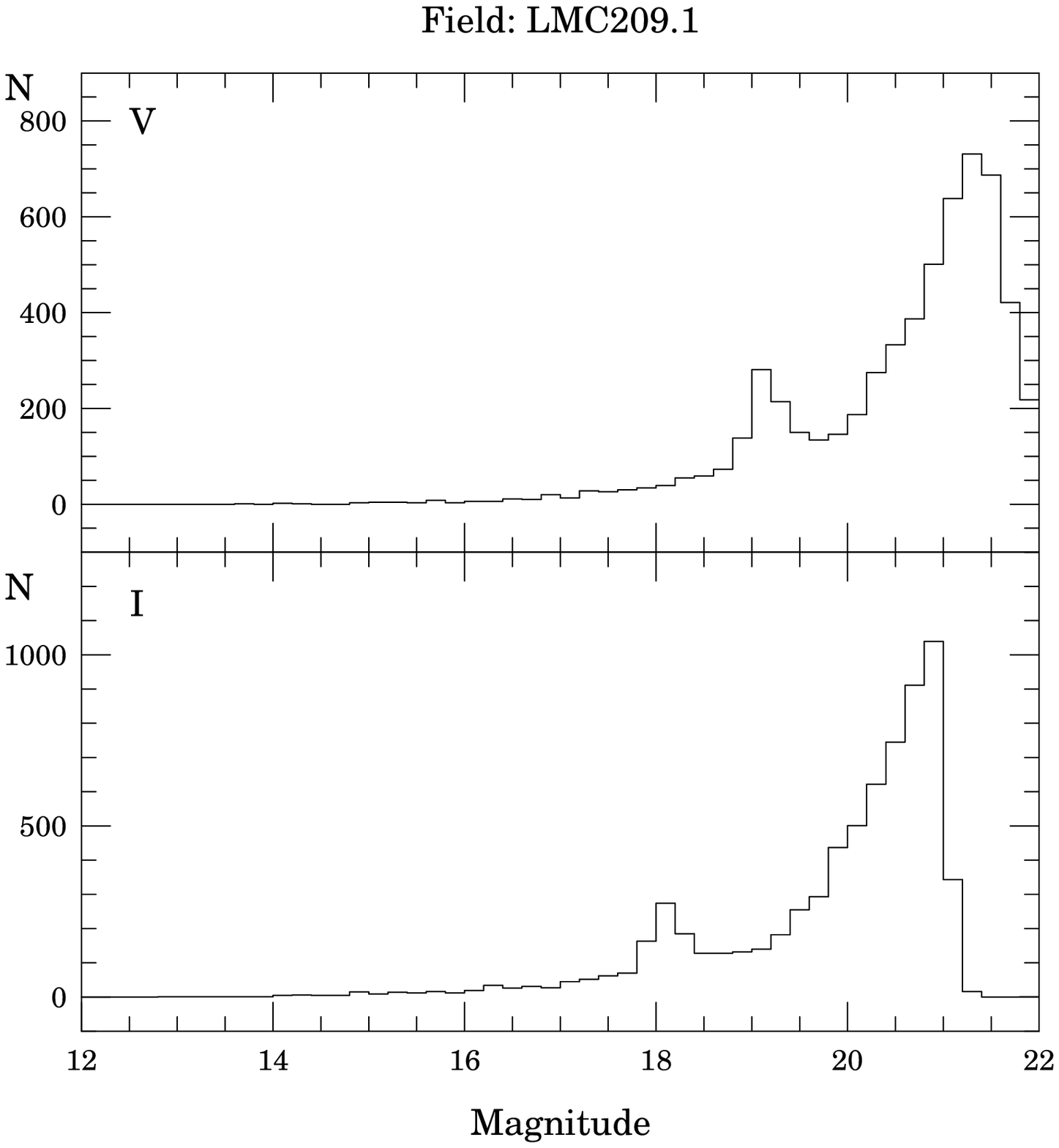}}
\FigCap{Same as in Fig.~5 for the outer subfield LMC209.1.} 
\end{figure}
Figs.~4 and 5 show the histograms of objects in the {\it V} and {\it
I}-band for the same two fields. One can notice that the completeness of
the maps is high and reaches $V\approx20.5{-}21$~mag and
$I\approx20{-}20.5$~mag. Again, as one can expected, completeness is a
function of stellar crowding.

\begin{landscape}
\renewcommand{\arraystretch}{1}
\MakeTableSep{
r@{\hspace{21pt}}
c@{\hspace{21pt}}
c@{\hspace{21pt}}
r@{\hspace{21pt}}
c@{\hspace{21pt}}
c@{\hspace{21pt}}
r@{\hspace{21pt}}
c@{\hspace{21pt}}
c@{\hspace{21pt}}
r@{\hspace{21pt}}
c@{\hspace{21pt}}
r@{\hspace{21pt}}
c@{\hspace{21pt}}
c}{12.5cm}
{OGLE-III Photometric Map of the LMC100.1 subfield.}
{\hline
\noalign{\vskip4pt}
ID & RA    & DEC   & $X$~~~~ & $Y$ & $V$ & $V-I$ & $I$ & $N_V$ & $N^{\rm bad}_V$ & $\sigma_V$ & $N_I$ & $N^{\rm bad}_I$ & $\sigma_I$\\
   &(2000) & (2000)&&&&&&&&&&&\\
\noalign{\vskip4pt}
\hline
\noalign{\vskip4pt}
1 & 5\uph19\upm05\zdot\ups97 & $-69\arcd33\arcm17\zdot\arcs4$ &  57.08 &~~56.26 & 16.647 & $ 2.483$ & 14.164 & 35 & $ 0$ & 0.127 &   6 & 0 & 0.058\\
2 & 5\uph19\upm08\zdot\ups62 & $-69\arcd32\arcm15\zdot\arcs3$ & 296.00 & 110.54 & 16.331 & $ 1.922$ & 14.409 & 67 & $-1$ & 0.208 & 399 & 0 & 0.077\\
3 & 5\uph19\upm08\zdot\ups67 & $-69\arcd30\arcm56\zdot\arcs1$ & 600.20 & 112.84 & 17.715 & $ 3.062$ & 14.652 & 47 & $ 0$ & 0.157 & 399 & 0 & 0.065\\
4 & 5\uph19\upm09\zdot\ups13 & $-69\arcd30\arcm39\zdot\arcs7$ & 663.16 & 122.23 & 16.318 & $ 1.847$ & 14.471 & 48 & $ 0$ & 0.018 & 402 & 0 & 0.010\\
5 & 5\uph19\upm10\zdot\ups04 & $-69\arcd29\arcm46\zdot\arcs7$ & 866.65 & 141.47 & 14.495 & $ 0.354$ & 14.141 & 48 & $ 0$ & 0.005 & 402 & 0 & 0.007\\
6 & 5\uph19\upm11\zdot\ups95 & $-69\arcd32\arcm00\zdot\arcs9$ & 351.08 & 177.73 & 17.470 & $ 3.182$ & 14.288 & 44 & $ 0$ & 0.680 & 404 & 0 & 0.214\\
7 & 5\uph19\upm13\zdot\ups76 & $-69\arcd30\arcm00\zdot\arcs3$ & 814.39 & 216.32 & 16.530 & $ 2.092$ & 14.438 & 48 & $ 0$ & 0.029 & 404 & 0 & 0.015\\
8 & 5\uph19\upm14\zdot\ups12 & $-69\arcd29\arcm03\zdot\arcs2$ &1033.50 & 224.43 & 14.141 & $ 0.188$ & 13.953 & 48 & $ 0$ & 0.004 & 404 & 0 & 0.006\\
9 & 5\uph19\upm14\zdot\ups53 & $-69\arcd32\arcm48\zdot\arcs2$ & 168.82 & 228.84 & 16.917 & $ 2.326$ & 14.591 & 87 & $-1$ & 0.072 & 385 & 0 & 0.033\\
10& 5\uph19\upm16\zdot\ups07 & $-69\arcd30\arcm03\zdot\arcs1$ & 803.47 & 262.79 & 14.739 & $ 0.780$ & 13.959 & 48 & $ 0$ & 0.005 & 404 & 0 & 0.005\\
11& 5\uph19\upm17\zdot\ups31 & $-69\arcd31\arcm47\zdot\arcs3$ & 402.93 & 285.85 & 15.366 & $ 1.367$ & 13.998 & 48 & $ 0$ & 0.006 & 404 & 0 & 0.005\\
12& 5\uph19\upm17\zdot\ups46 & $-69\arcd29\arcm50\zdot\arcs9$ & 850.10 & 290.92 & 16.813 & $ 2.172$ & 14.641 & 48 & $ 0$ & 0.072 & 404 & 0 & 0.031\\
13& 5\uph19\upm19\zdot\ups14 & $-69\arcd30\arcm12\zdot\arcs7$ & 766.32 & 324.45 & 17.416 & $ 3.007$ & 14.408 & 47 & $ 0$ & 0.286 & 404 & 0 & 0.189\\
14& 5\uph19\upm19\zdot\ups61 & $-69\arcd32\arcm51\zdot\arcs7$ & 155.22 & 330.93 & 16.432 & $ 2.245$ & 14.187 & 84 & $-1$ & 0.048 & 366 & 0 & 0.021\\
15& 5\uph19\upm19\zdot\ups54 & $-69\arcd31\arcm44\zdot\arcs6$ & 413.08 & 330.83 & 16.979 & $ 2.371$ & 14.608 & 48 & $ 0$ & 0.131 & 404 & 0 & 0.062\\
16& 5\uph19\upm23\zdot\ups42 & $-69\arcd33\arcm20\zdot\arcs3$ &  44.83 & 407.14 & 16.310 & $ 9.999$ & 99.999 & 40 & $ 0$ & 0.052 &   0 & 0 & 9.999\\
17& 5\uph19\upm25\zdot\ups72 & $-69\arcd29\arcm44\zdot\arcs3$ & 874.65 & 457.67 & 15.044 & $ 1.200$ & 13.844 & 48 & $ 0$ & 0.006 & 404 & 0 & 0.004\\
18& 5\uph19\upm27\zdot\ups42 & $-69\arcd31\arcm18\zdot\arcs7$ & 511.94 & 489.95 & 15.476 & $ 1.214$ & 14.262 & 48 & $ 0$ & 0.008 & 404 & 0 & 0.005\\
19& 5\uph19\upm27\zdot\ups78 & $-69\arcd30\arcm30\zdot\arcs2$ & 698.01 & 498.27 & 14.010 & $ 9.999$ & 99.999 & 48 & $ 0$ & 0.153 &   0 & 0 & 9.999\\
20& 5\uph19\upm29\zdot\ups18 & $-69\arcd30\arcm57\zdot\arcs9$ & 591.71 & 525.78 & 13.920 & $-0.135$ & 14.055 & 48 & $ 0$ & 0.006 & 404 & 0 & 0.008\\
21& 5\uph19\upm32\zdot\ups47 & $-69\arcd30\arcm34\zdot\arcs9$ & 679.44 & 592.49 & 15.383 & $ 9.999$ & 99.999 & 48 & $ 0$ & 0.061 &   0 & 0 & 9.999\\
22& 5\uph19\upm32\zdot\ups64 & $-69\arcd30\arcm32\zdot\arcs9$ & 687.26 & 596.09 & 16.676 & $ 9.999$ & 99.999 & 48 & $ 0$ & 0.026 &   0 & 0 & 9.999\\
23& 5\uph19\upm33\zdot\ups45 & $-69\arcd30\arcm18\zdot\arcs0$ & 744.46 & 612.71 & 14.765 & $ 1.573$ & 13.192 & 48 & $ 0$ & 0.008 & 404 & 0 & 0.006\\
24& 5\uph19\upm33\zdot\ups84 & $-69\arcd30\arcm18\zdot\arcs0$ & 744.28 & 620.53 & 14.906 & $ 1.322$ & 13.584 & 48 & $ 0$ & 0.008 & 404 & 0 & 0.007\\
25& 5\uph19\upm37\zdot\ups93 & $-69\arcd31\arcm46\zdot\arcs7$ & 402.99 & 700.88 & 15.501 & $ 2.142$ & 13.359 & 48 & $ 0$ & 0.046 & 404 & 0 & 0.017\\
\noalign{\vskip4pt}
\hline}
\end{landscape}

Figs.~6--9 show several color--magnitude diagrams (CMDs) constructed
directly from the OGLE-III maps. They include CMD of one of the densest
central fields of the LMC: LMC100.1, high extinction region near the
Tarantula nebula: LMC175.6, as well as regions in the outskirts of this
galaxy (LMC121.1 in the western wing and LMC186.1 in the eastern wing).
Figs.~6--9 clearly illustrate the quality of data and reveal in details
the characteristic features of the LMC stellar populations, as well as
show how interstellar extinction affects the CMDs.

\Section{Data Availability}
The OGLE-III Photometric Maps of the LMC are available to the astronomical
community from the OGLE Internet Archive:

\centerline{{\it http://ogle.astrouw.edu.pl \hfil  ftp://ftp.astrouw.edu.pl/ogle3/maps/lmc/}}

Beside tables with photometric data and astrometry for each of the
subfields the {\it I}-band reference images are also included. Usage of
the data is fully allowed, requiring only  the proper acknowledgment to
the OGLE project.

\Acknow{This paper was partially supported by the Polish MNiSW grant
N20303032/4275 to AU and NN203293533 to IS and  by the
Foundation for Polish Science through the Homing Program.}

\newpage
\centerline{Captions of JPEG figures.}
\vskip15pt
\noindent
Fig.~1. OGLE-III fields in the LMC (black squares: 100--215) overplotted
on the image obtained by the ASAS all sky survey. Red strips (1--21)
mark OGLE-II fields. 
\vskip15pt
\noindent
Fig.~2. Standard deviation of magnitudes as a function of magnitude for
the central bar subfield LMC100.1.
\vskip15pt
\noindent
Fig.~3. Same as in Fig.~2 for the outer subfield LMC209.1
\vskip15pt
\noindent
Fig.~6. Color--magnitude diagram for the central bar subfield LMC100.1.
\vskip15pt
\noindent
Fig.~7. Color--magnitude diagram for the variable extinction subfield LMC175.6.
\vskip15pt
\noindent
Fig.~8. Color--magnitude diagram for the western wing subfield LMC121.1
\vskip15pt
\noindent
Fig.~9. Color--magnitude diagram for the eastern wing subfield LMC186.1

\end{document}